\documentclass[12pt]{article}
\usepackage{a4}
\usepackage{cite}
\usepackage{pslatex}
\usepackage{graphicx}
\usepackage[latin1]{inputenc}
\usepackage[T1]{fontenc}

\def\sign(#1){(\!-\!1)^{#1}}
\def\binom(#1,#2){ (\!\!
         \begin{array}{c} #1 \\ #2 \end{array}\!\! ) }

\def\keyword#1{{\bfseries\sffamily #1}}

\def\XSummer{{\sffamily\scshape XSummer}}
\def\Summer{{\sffamily\scshape Summer}}
\def\Form{{\sffamily\scshape Form}}
\def\Maple{{\sffamily\scshape Maple}}
\def\Mathematica{{\sffamily\scshape Mathematica}}

\makeatletter
\def\@sect#1#2#3#4#5#6[#7]#8{\ifnum #2>\c@secnumdepth
  \def\@svsec{}\else 
  \refstepcounter{#1}\edef\@svsec{\csname the#1\endcsname.\hskip0.5em}\fi
  \@tempskipa #5\relax
  \ifdim \@tempskipa>\z@
    \begingroup 
      #6\relax
      \@hangfrom{\hskip #3\relax\@svsec}{\interlinepenalty \@M #8\par}%
    \endgroup
    \csname #1mark\endcsname{#7}\addcontentsline
      {toc}{#1}{\ifnum #2>\c@secnumdepth \else
        \protect\numberline{\csname the#1\endcsname}\fi #7}%
  \else
    \def\@svsechd{#6\hskip #3\@svsec #8\csname #1mark\endcsname
      {#7}\addcontentsline{toc}{#1}{\ifnum #2>\c@secnumdepth \else
        \protect\numberline{\csname the#1\endcsname}\fi #7}}%
  \fi \@xsect{#5}}
\@addtoreset{equation}{section}
\@addtoreset{figure}{section}
\makeatother

\renewcommand\theequation{\ifnum \value{section}>0
 \arabic{section}.\arabic{equation}%
\else
\arabic{equation}%
\fi}
\renewcommand\thefigure{\ifnum \value{section}>0
 \arabic{section}.\arabic{figure}%
\else
\arabic{figure}%
\fi}

\begin{document}

\begin{titlepage}
\noindent
DESY 05-104 \hfill {\tt math-ph/0508008}\\
SFB/CPP-05-24 \\
CERN-PH-TH/2005-124 \\
\vspace{1.3cm}
\begin{center}
\Large{\bf 
-- \XSummer\  --\\ 
Transcendental Functions and Symbolic Summation in \Form}\\
\vspace{1.5cm}
\large
S. Moch$^{\, a}$ and P. Uwer$^{\, b}$\\
\vspace{1.2cm}
\normalsize
{\it $^a$Deutsches Elektronensynchrotron DESY \\
\vspace{0.1cm}
Platanenallee 6, D--15735 Zeuthen, Germany}\\
\vspace{0.5cm}
{\it $^b$Department of Physics, TH Division, CERN \\
\vspace{0.1cm}
CH-1211 Geneva 23, Switzerland}\\
\vspace{1.4cm}

\centerline{July 2005} 

\vspace{1.4cm}
\large
{\bf Abstract}
\vspace{-0.2cm}
\end{center}
Harmonic sums and their generalizations are 
extremely useful in the evaluation of higher-order perturbative corrections 
in quantum field theory. Of particular interest have been the 
so-called nested sums,
where the harmonic sums and their generalizations appear as
building blocks, originating for example from the expansion of generalized 
hypergeometric functions around integer values of the parameters. 
In this Letter we discuss the implementation of several algorithms to solve 
these sums by algebraic means, using the computer algebra system \Form.

\vfill
\end{titlepage}

\section*{Program summary}

{\it Title of program:} \XSummer \\[2mm]
{\it Version:} 1.0 \\[2mm]
{\it Catalogue number:} \\[2mm]
{\it Program summary URL:} {\tt http://www-zeuthen.desy.de/\~{}moch/xsummer}\\[2mm]
{\it E-mail:} {\tt sven-olaf.moch@desy.de}, {\tt peter.uwer@cern.ch} \\[2mm]
{\it License:} GNU Public License and \Form\ License \\[2mm]
{\it Computers:} all \\[2mm]
{\it Operating system:} all\\[2mm]
{\it Program language:} \Form \\[2mm]
{\it Memory required to execute:} Depending on the complexity of the problem, 
        recommended at least 64 MB RAM.\\[2mm]
{\it Other programs called:} none\\[2mm]
{\it External files needed:} none\\[2mm]
{\it Keywords:} Symbolic summation, Multiple polylogarithms, Transcendental functions.\\[2mm]
{\it Nature of the physical problem: }
         Systematic expansion of higher transcendental functions in
         a small parameter.
         The expansions arise in the calculation of loop integrals
         in perturbative quantum field theory.\\[2mm]
{\it Method of solution:} Algebraic manipulations of nested sums.\\[2mm]
{\it Restrictions on complexity of the problem:} Usually limited only by the available disk space.\\[2mm]
{\it Typical running time:} Dependent on the complexity of the problem.
\newpage

%
% ---------------------------------------------------------------------
%
\setcounter{equation}{0}
\section{Introduction}
\label{sec:introduction}
%
% ---------------------------------------------------------------------
%
Symbolic summation amounts to finding a closed-form expression for 
a given sum or series.
Systematic studies have been pioneered by Euler~\cite{Euler:1775xx}, and
for specific sums, exact formulae have been known for a long time, series representations 
of transcendental functions being a prominent example.
Today, general classes of sums, for example so-called harmonic sums, 
have been investigated (see e.g. Refs.~\cite{Hoffman:refs})
and symbolic summation has further advanced through the development 
of algorithms suitable for computer algebra systems.
Here, the possibility to obtain exact solutions 
by means of recursive methods has lead to significant progress, 
for instance in the summation of rational or hypergeometric series, 
see e.g. Ref.~\cite{Petkovsek:1997ab}.

In quantum field theory, higher-order corrections in perturbation theory 
require the evaluation of so-called Feynman diagrams, which describe 
real and virtual particles in a given scattering process.
In mathematical terms, Feynman diagrams are given as integrals over 
the loop momenta of the associated particle propagators. 
These integrals may depend on multiple scales and are usually divergent, 
thus requiring some regularization.
The standard choice is dimensional regularization, i.e. 
an analytical continuation of the dimensions of space-time from 4 to $D$, 
which keeps underlying gauge symmetries manifestly invariant.
Analytical expressions for Feynman integrals in $D$ dimensions may lead to transcendental 
or generalized hypergeometric functions, which have a series representation through nested sums 
with symbolic arguments. 
The main computational task is then to obtain the Laurent series upon expansion of the relevant 
functions in the small parameter $\epsilon=(D-4)/2$.

It is the aim of the present Letter to discuss the implementation of several algorithms~\cite{Moch:2001zr}
for these tasks in the computer algebra system \Form\ ~\cite{Vermaseren:2000nd,Vermaseren:2002rp}.
The resulting package \XSummer\  has already been used in full-fledged calculations
in particle physics, for instance in calculating higher-order perturbative corrections 
in Quantum Chromodynamics, see e.g. Ref.~\cite{Moch:2002hm}.
We hope that it may also be useful for a larger community, as it exceeds current built-in
capabilities of commercial computer algebra systems such as \Maple\  or \Mathematica\  
in the expansion of (generalized) hypergeometric functions.

To give a concrete example for the kind of problems we aim at, 
consider the hypergeometric function $_2F_1$, which has a series
representation for arguments $|x| \le 1$:
\begin{eqnarray}
\label{eq:ex-2F1}
&&_2F_1(a\epsilon,b\epsilon,1-c\epsilon,x) 
= 
\sum\limits_{j=0}^{\infty} 
{(a\epsilon)^{\overline{j}} (b\epsilon)^{\overline{j}} x^j \over (1-c\epsilon)^{\overline{j}} j!} \nonumber
\\ 
&=&
1 + ab {\rm Li}_2(x) \epsilon^2
+ ab \bigl\{ c {\rm Li}_3(x) + (a+b+c) {\rm S}_{1,2}(x) \bigr\} \epsilon^3
+ {\cal O}(\epsilon^4)\, ,
\end{eqnarray}
where the $(a\epsilon)^{\overline{j}} = \Gamma(j+a\epsilon)/\Gamma(a\epsilon)$
are so-called rising factorials (in the literature also known as Pochhammer symbols).
Here, we have expressed the coefficients of the Laurent series in $\epsilon$ through 
standard polylogarithms ${\rm Li}_n$ and Nielsen functions ${\rm S}_{n,p}$, see e.g. 
Ref.~\cite{lewin:book}.

Our choice of \Form\  for the implementation is based on two reasons, 
one being that \Form\  is extremely fast and flexible when dealing with large expressions.
\Form\  allows for a very compact notation and is equipped 
with a pattern matcher well suited to solve our problem.
The ability to handle large-size expressions (of the order of the computer memory) 
is crucial, as they generally occur 
at intermediate stages in the quantum field theory calculations mentioned above, 
for instance when the Laurent expansions in $\epsilon$ have to be done to very
high order.

The other main motivation for choosing \Form\  is due to the existing \Summer\  
package~\cite{Vermaseren:1998uu}. The \Summer\  package  in \Form\  is capable 
of solving nested sums in terms of harmonic sums, a feature that has also 
been used extensively in recent cutting-edge calculations of structure functions 
to three loops in Quantum Chromodynamics~\cite{Moch:2004pa,Vogt:2004mw,Vermaseren:2005qc}.
Here, the \XSummer\  package, being capable of handling (multiple) scales,  e.g. 
in Eq.~(\ref{eq:ex-2F1}), provides the obvious extension.
In the development of \XSummer, we have also benefited from the fact that some parts 
of the underlying algorithmic structures could be literally taken from \Summer.

A major disadvantage of using \Form\  for \XSummer\  is certainly the lack of internal 
algorithms for particular operations on polynomials, such as factorization.
We will comment on that in the text.
Also, we note that an implementation of the algorithms of Ref.~\cite{Moch:2001zr} 
within the GiNaC framework~\cite{Bauer:2000cp} is available~\cite{Weinzierl:2002hv} .

The outline of the Letter is as follows. In Section~\ref{sec:harmonicsums}, 
we briefly give the basics of generalized sums and recall the algorithms of Ref.~\cite{Moch:2001zr}.
In Section~\ref{sec:xsummer}, we present the \XSummer\  package and discuss details of the implementation.
Section~\ref{sec:examples} features an extensive set of tests with various sample calculations,
including e.g. Ref.~\cite{Moch:2002hm}. We conclude in Section~\ref{sec:conclusion}. 

%
% ---------------------------------------------------------------------
%
\setcounter{equation}{0}
\section{Harmonic sums and their generalizations}
\label{sec:harmonicsums}
%
% ---------------------------------------------------------------------
%
The basic recursive definition of $S$-sums is given by~\cite{Moch:2001zr}
\begin{eqnarray}
\label{eq:def-S}
S(n) & = & \left\{ \begin{array}{cc}
1, & n > 0, \\
0, & n \le 0, \\
\end{array} \right.
\nonumber \\
S(n;m_1,...,m_k;x_1,...,x_k) 
&=& \sum\limits_{i=1}^n \frac{x_1^i}{i^{m_1}} S(i;m_2,...,m_k;x_2,...,x_k)\, .
\end{eqnarray}
These are the basic objects that we will manipulate in the following.
Generally, we have all $|x_i| \le 1$ in Eq.~(\ref{eq:def-S}). 
The sum of all $m_i$ is called the weight of the sum, while the index $k$ denotes the depth. 
This definition actually includes as special cases the series representations of classical polylogarithms, 
Nielsen functions, as well as multiple and harmonic polylogarithms~\cite{Goncharov,Borwein,Remiddi:1999ew}.
For all $x_i = 1$, the above definition reduces to harmonic
sums~\cite{Euler:1775xx,Vermaseren:1998uu,Blumlein:2003gb} and, if additionally the upper 
summation boundary $n \to \infty$, one recovers the (multiple) zeta values 
associated to Riemann's zeta-function~\cite{Hoffman:refs}.

An equivalent representation of $S$-sums reads
\begin{eqnarray}
S(n;m_1,...,m_k;x_1,...,x_k)  
&=& 
\sum\limits_{n\ge i_1 \ge i_2\ge \ldots\ge i_k \ge 1}
\frac{x_1^{i_1}}{{i_1}^{m_1}}\ldots \frac{x_k^{i_k}}{{i_k}^{m_k}}\, .
\end{eqnarray}
We note that $S$-sums are closely related to so-called $Z$-sums, the difference being the upper summation boundary
for the nested sums: $(i_k-1)$ for $Z$-sums, $i_k$ for $S$-sums, see Ref.~\cite{Moch:2001zr}.
One can algebraically convert $Z$-sums to $S$-sums and vice versa~\cite{Hoffman1,Hoffman2}. 
We rely entirely on $S$-sums in our discussions, but nevertheless provide 
the procedure \keyword{ConvStoZ} since, in some cases, $Z$-sums may be more favourable.

The $S$-sums obey the well-known algebra of multiplication, 
a straightforward generalization of the
results on the multiplication of harmonic sums~\cite{Vermaseren:1998uu}.
The basic formula reads
\begin{eqnarray}
\label{eq:prod-S}
\lefteqn{
S(n;m_1,...,m_k;x_1,...,x_k) \times S(n;m_1^\prime,...,m_l^\prime;x_1^\prime,...,x_l^\prime) } 
& & \nonumber \\
&=& 
\sum\limits_{i_1=1}^n \frac{x_1^{i_1}}{i_1^{m_1}}
S(i_1;m_2,...,m_k;x_2,...,x_k) S(i_1;m_1^\prime,...,m_l^\prime;x_1^\prime,...,x_l^\prime) 
\nonumber \\
&  & 
+ \sum\limits_{i_2=1}^n \frac{{x_1^\prime}^{i_2}}{i_2^{m_1^\prime}}
S(i_2;m_1,...,m_k;x_1,...,x_k) S(i_2;m_2^\prime,...,m_l^\prime;x_2^\prime,...,x_l^\prime) 
\nonumber \\
&  & 
- \sum\limits_{i=1}^n \frac{\left(x_1 x_1^\prime \right)^{i}}{i^{m_1+m_1^\prime}}
S(i;m_2,...,m_k;x_2,...,x_k) S(i;m_2^\prime,...,m_l^\prime;x_2^\prime,...,x_l^\prime)\, ,
\end{eqnarray}
which works recursively in the depth of the individual sums.
The algorithm allows the expression of any product of nested sums  as a sum of single
nested sums, hence in a canonical form, which is an important feature 
for practical applications.
The underlying algebraic structure in Eq.~(\ref{eq:prod-S}) is a Hopf algebra,
being realized as a quasi-shuffle algebra here, see e.g. Refs.~\cite{Hoffman:refs,Moch:2001zr} .
The algorithm  can be implemented very efficiently on a computer, 
see the procedure \keyword{BasisS} in Section~\ref{sec:xsummer}.

In our applications, such as in Eq.~(\ref{eq:ex-2F1}), we encounter Gamma-functions, 
which we have to expand in the small parameter $\epsilon$ before the actual manipulation of the nested sums. 
This proceeds according to the well-known formula for the expansion of the Gamma-function around positive integer 
values,
\begin{eqnarray}
\label{eq:gamma_integer_pos}
\frac{\Gamma(n+1+\epsilon)}{\Gamma(1+\epsilon)} 
&=& 
\Gamma(n+1)\, \exp \left( - \sum\limits_{k=1}^\infty \epsilon^k\, \frac{\sign(k)}{k}\, S_k(n) \right)\, .
\end{eqnarray}
Similarly, the expansion of the Gamma-function around negative integer values can be reduced to the case 
in Eq.~(\ref{eq:gamma_integer_pos}) with the help of the following relation 
(e.g. Ref.~\cite{Erdelyi} p.~3),
\begin{eqnarray}
\label{eq:gamma_pos_neg}
\frac{\Gamma(-n+1+\epsilon)}{\Gamma(1+\epsilon)} 
&=&
\sign(n)\, \frac{\Gamma(-\epsilon)}{\Gamma(n-\epsilon)}\, ,
\end{eqnarray}
which yields
\begin{eqnarray}
\label{eq:gamma_integer_neg}
\frac{\Gamma(-n+1+\epsilon)}{\Gamma(1+\epsilon)} 
&=&
\frac{1}{\epsilon}\, \frac{\sign(n-1)}{\Gamma(n)}\, 
\exp \left( \sum\limits_{k=1}^\infty \epsilon^k \frac{1}{k} S_k(n-1) \right)\, .
\end{eqnarray}

\subsection{Algorithms}
For the manipulation of the $S$-sums, we classify four types of transcendental sums. 
These types of sums are dealt with in the algorithms A to D given below.
All sums in these classes can be solved recursively, i.e. they can be expressed in canonical form.
The underlying algorithms realize a creative telescoping. 
They either reduce successively the depth or the weight of the inner sum, 
so that eventually the inner nestings vanish and the results 
can be written in the basis of Eq.~(\ref{eq:def-S}).
The procedure generally relies on algebraic manipulations, such as partial fractioning of denominators, 
shifts of the summation ranges and synchronization of summation boundaries of the individual sums.
Another crucial ingredient is, of course, the quasi-shuffle algebra of multiplication in Eq.~(\ref{eq:prod-S}).

\subsubsection*{Basic definition (type A):}
Here we consider sums over $i$ involving only $S(i;...)$, of the form, 
\begin{eqnarray}
\label{eq:type-A}
S(n;m_1,...,m_k;x_1,...,x_k) &=&
        \sum\limits_{i=1}^n {x_1^i \over (i+a)^{m_1}} S(i+b;m_2,...,m_k;x_2,...,x_k) \, ,
\end{eqnarray}
where we assume that $a,b $ are (non-symbolic) integers. The upper summation limit is allowed to be infinity.

\subsubsection*{Convolution (type B):}
Here we consider convolutions, i.e. sums over $i$ involving both $S(i;...)$ and $S(n-i;...)$, of the form,
\begin{eqnarray}
\label{eq:type-B}
  &&\sum\limits_{i=1}^{n-1} \bigg( 
  {x_1^i \over (i+a)^{m_1}} S(i+b;m_2,...,m_k;x_2,...,x_k) \nonumber \\
  &&\hspace{0.6cm}\times {(x^{\prime}_1)^{n-i} \over (n-i+a^\prime)^{m^{\prime}_1}} 
  S(n-i+b^\prime;m^{\prime}_2,...,m^{\prime}_l;x^{\prime}_2,...,x^{\prime}_l)
  \bigg)\, ,
\end{eqnarray}
where all $a, a^\prime, b$  and $b^\prime$ must be (non-symbolic) integers. Note that the upper summation limit 
is $(n-1)$ and thus consistent with the defining range of the $S$-sums.

\subsubsection*{Conjugation  (type C):}
Here we consider conjugations, i.e. sums over $i$ involving 
\begin{displaymath}
  (-1)^i\, S(i\,;m_1,\ldots;x_1,\ldots) 
\end{displaymath}
and a binomial 
\begin{displaymath}
  \left(\begin{array}{c} n \\ i \end{array} \right), 
\end{displaymath}
of the form,
\begin{eqnarray}
\label{eq:type-C}
  - \sum\limits_{i=1}^{n}\, \left( \begin{array}{c} n \\ i \end{array} \right)\, 
  (-1)^i {x_1^i \over (i+a)^{m_1}} S(i+b;m_2,...,m_k;x_2,...,x_k)\, ,
\end{eqnarray}
where $a, b$ are  (non-symbolic) integers.
The upper summation limit should not be infinity. Again, the upper summation limit 
$n$ is consistent with the defining range of the binomial.
Sums of this type cannot be reduced to $S$-sums with upper summation limit $n$ alone.
However, they can be reduced to $S$-sums with upper summation limit $n$ and
multiple polylogarithms (which are $S$-sums to infinity).

\subsubsection*{Binomial convolution (type D):}
Here we consider binomial convolutions, i.e. 
sums over $i$ involving $(-1)^i\, S(i;...)$, $S(n-i;...)$ and 
a binomial, of the form,
\begin{eqnarray}
\label{eq:type-D}
{\lefteqn{
  - \sum\limits_{i=1}^{n-1}\, \left(\begin{array}{c} n \\ i \end{array} \right)\, 
  (-1)^i {x_1^i \over (i+a)^{m_1}} S(i+b;m_2,...,m_k;x_2,...,x_k) }}
\nonumber\\
& &
\times 
{(x^\prime_1)^{n-i} \over (n-i+a^\prime)^{m^{\prime}_1}} S(n-i+b^\prime;m^{\prime}_2,...,m^{\prime}_l;x^{\prime}_2,...,x^{\prime}_l)\, .
\end{eqnarray}
Here, all $a, a^\prime, b$  and $b^\prime$ must be (non-symbolic) integers.
Yet again, the upper summation limit $(n-1)$ reflects the defining range of the binomial and the $S$-sums.
As for sums of type C, we cannot relate them  to
$S$-sums with upper summation limit $(n-1)$ alone, but
we can reduce them to $S$-sums with upper summation limit $(n-1)$ and
multiple polylogarithms (which are $S$-sums to infinity).

%
% ---------------------------------------------------------------------
%
\setcounter{equation}{0}
\section{The \XSummer\ package}
\label{sec:xsummer}
%
% ---------------------------------------------------------------------
%
In this section we give a short description of the \XSummer\ package. 
In particular we explain the notation that we use in the package
and the main routines that act as a front-end to a bunch of smaller 
routines used internally. 
At the end of this section we also comment briefly on the internal routines, 
although the user might not want to call them directly.

\subsection{Basic syntax --- form follows function}
The \XSummer\ package is written using the computer algebra system
\Form. Unlike programs such as \Maple\ or \Mathematica\  
the program \Form\ provides only a very limited set of built-in capabilities. 
\Form\ is mainly a highly efficient pattern matcher. 
In choosing a syntax/notation for the \XSummer\ package it is therefore important 
to ensure that all basic algorithms can be implemented as simple pattern matching 
and, at the same time, make this pattern matching as simple as possible.

Generically, we use the function \keyword{R} with an arbitrary number of arguments
to denote a list of integer parameters. 
Similarly, \keyword{X} is used to denote a list of symbolic arguments. 
Due to the internal limitations of \Form\ with respect to polynomial algebra,
it is of some advantage to define, for various functions, a distinct multiplicative inverse, 
which helps in bringing expressions to a normal form. 
Examples are the pairs \keyword{Gamma} and \keyword{InvGamma}.
Also note, that the summation symbol simply appears as a function multiplying 
all terms that should be summed over.
For instance, a sum of the form shown in Eq.~(\ref{eq:type-C}),
\begin{equation}
  \label{eq:AlgCDemo}
  \sum\limits_{j_1=1}^{n}\, \left( {n \atop j_1} \right)\, 
  (-1)^{j_1} {x_1^{j_1} \over (j_1+3)^{2}} 
  S(j_1+2;4,7,1,1;x_2,...,x_5)\, ,
\end{equation}
would be written as:
\begin{verbatim}
   sum(j1,1,n) * bino(n,j1) *  pow(-x1,j1) * den(j1+3)^2 *
      S(R(4,7,1,1),X(x2,x3,x4,x5),j1+2);
\end{verbatim}
Table~\ref{tab:Keywords} shows the complete set of keywords for the \XSummer\ package. 
\begin{table}[htbp] 
    \label{tab:Keywords}
  \renewcommand{\arraystretch}{1.25}
  \begin{center}
    \leavevmode
    \begin{tabular}[htbp]{lll}
      Name & Description & Example \\ \hline
      \keyword{bino} & binomial coefficient &
      \keyword{bino(n,i)} $\to \left({n \atop i}\right)$ \\
      \keyword{delta} & Kronecker delta &
      \keyword{delta(x)} 
      $\to \left\{{1,\,\, x=0\atop 0,\,\, x\not = 0}\right.$ \\
      \keyword{deltap} & inverse Kronecker delta &
      \keyword{deltap(x)} 
      $\to (1-\mbox{\keyword{delta}}(x))$ \\
      \keyword{den} & denominator function & \keyword{den}(x) $\to
      {1\over x}$ \\  
      \keyword{ep} & expansion parameter $\epsilon$ &
      \keyword{ep} 
      $\to \epsilon$ \\
      \keyword{epow} & powers of epsilon & \keyword{epow}(n) $\to
      \epsilon^n$ \\ 
      \keyword{fac} & factorial function& \keyword{fac}(n) $\to
      {n!}$ \\ 
      \keyword{inf} & parameter for $\infty$ &
      \keyword{inf} 
      $\to \infty$ \\
      \keyword{invfac} & inverse factorial function& 
      \keyword{invfac}(n) $\to
      {1\over n!}$ \\ 
      \keyword{num} & numerator function & \keyword{num}(x) $\to
      {x}$ \\   
      \keyword{pow} & power function & \keyword{pow}(x,a) $\to
      x^a$ \\ 
      \keyword{sign} & sign function &
      \keyword{sign(n)} 
      $\to (-1)^n$ \\
      \keyword{sum} & summation symbol &
      \keyword{sum(j,i1,i2)} $\to \sum_{j=i_1}^{i_2}$ \\
      \keyword{theta} & theta function &
      \keyword{theta(x)} 
      $\to \left\{{1,\,\, x\ge 0\atop 0,\,\, x< 0}\right.$\\
      \keyword{z2, z3,\ldots} & values of the zeta-function &
      \keyword{z2}$\to {\zeta(2)},\ldots$ \\
      \keyword{Gamma} & Gamma-function & \keyword{Gamma}(x) $\to
      {\Gamma(x)}$ \\ 
      \keyword{InvGamma} & inverse Gamma-function & 
      \keyword{InvGamma}(x) $\to
      {1\over \Gamma(x)}$ \\ 
      \keyword{S(R($m_1,\ldots$),X($x_1,\ldots$),n)} & 
      $S$-sum& \keyword{S(R($m_1,\ldots,$),X($x_1,\ldots$),n)}\\
      &&$\to
      S(n; m_1,\ldots,m_k; x_1,\ldots,x_k)$ \\ 
      \keyword{Z(R($m_1,\ldots$),X($x_1,\ldots$),n)} & 
      Z-sum& \keyword{Z(R($m_1,\ldots,$),X($x_1,\ldots$),n)}\\
      &&$\to
      Z(n; m_1,\ldots,m_k; x_1,\ldots,x_k)$ \\
      \hline
    \end{tabular}
    \caption{Basic objects appearing in  the input/output of the 
      \XSummer\ package}
  \end{center}
  \renewcommand{\arraystretch}{1.}
\end{table}
All the objects in Table~\ref{tab:Keywords} are defined in 
the file \keyword{declvars.h} which should be included when using
the \XSummer\ package.

Note that the \keyword{pow} function is reserved for symbolic variables to some power 
of a summation index. 
Writing 
\begin{verbatim}
   pow(j1+3,-2)
\end{verbatim}
instead of
\begin{verbatim}
   den(j1+3)^2
\end{verbatim}
would not work. 
This is just due to the way the pattern matching is realized in the package. 

Furthermore, it is assumed that the summation variables are always $j_1$, $j_2$, ... 
where the outermost sum runs over $j_1$, the next sum runs over $j_2$ and so on. 
The innermost sum is the sum over the highest $j_i$. 
Upon summation of a nested sum the innermost sum must be done first and we then 
work through to the outer sums. 
This is essentially done by calling the procedure \keyword{DoSum}, 
which takes as arguments the summation indexes of the innermost and the outermost sum. 
For example:
\begin{verbatim}
#call DoSum(3,1)
\end{verbatim}
would evaluate the sums containing $j_3$, $j_2$, $j_1$ (in this order).
Note that, when doing the sum over a specific $j_i$, all the objects
relevant for this sum are dressed internally with an additional index $i$ 
as part of the names of the symbols. 
For example the sum shown in Eq.~(\ref{eq:AlgCDemo}) would be converted internally to
\begin{verbatim}
sum1(j1,1,n) * bino1(n,j1) *  pow1(-x1,j1) * den1(j1+3)^2 *
    S1(R(4,7,1,1),X(x2,x3,x4,x5),j1+2);
\end{verbatim}
This is just to simplify the pattern matching. 
However, in the final result the {\it dressed objects} should never occur.
The following \Form\ script solves a much easier version of the example 
shown above (otherwise the result would be to lengthy to be reproduced here):
\begin{verbatim}
#define MAXSUM "1"
#define MAXWEIGHT "20"
#include declvars.h
nwrite stat;

L demo = sum(j1,1,n) * bino(n,j1) *  pow(-x1,j1) * den(j1+1) 
   * S(R(1,1),X(x2,x3),j1+1);

id bino(x1?,x2?) = fac(x1)*invfac(x2)*invfac(x1-x2); 

#call DoSum(1,1)

print;

.end;
\end{verbatim}
The result obtained from running \Form\ is given by:
\begin{verbatim}
demo =
   + acc(-1)*pow(1 - x1,1 + n)*den( - x1)*den(1 + n)
   *S(R(1,1),X(den(1 - x1) - den(1 - x1)*x1*x2,den(1 - x1*x2))
   ,1 + n)*theta( - 1 + n)+ acc(1)*pow(1 - x1,1 + n)*den( - x1)
   *den(1 + n)*S(R(1,1),X(den(1 - x1)- den(1 - x1)*x1*x2,
   den(1 - x1*x2) - den(1 - x1*x2)*x1*x2*x3),1 + n)*theta( - 1 + n)
    + acc(1)*den( - x1)*theta( - 1 + n)*x1*x2*x3
   ;
\end{verbatim}
The \keyword{acc} function is defined internally with the \Form\ 
declaration {\tt PolyFun} to collect similar objects together, 
in particular to accumulate powers of the expansion parameter $\epsilon$.
Its use is actually described in the \Form\ manual and the expanded 
result is simply obtained with
\begin{verbatim}
id acc(x?) = x;
\end{verbatim}

As mentioned earlier the integer parameters of the $S$-sums are
collected in \keyword{R}, while the symbolic arguments are collected
in \keyword{X}. 
Note that only basic simplifications are applied to the symbolic arguments. 
This is due to the fact that \Form\ does
not provide built-in routines to factorize or normalize expressions. 
Such procedures must be provided by the user and are highly dependent
on the problem studied. 
We will discuss this issue in Section~\ref{sec:examples}.

Our example, converted back to a more human readable notation is given by
\begin{eqnarray}
\label{eq:demoresult}
{\lefteqn{
\sum\limits_{j_1=1}^n \* \left( \begin{array}{c} n \\ j_1 \end{array} \right) 
\* {(-x_1)^{j_1} \over j_1+1} \* S(j_1+1;1,1;x_2,x_3) 
\, = \,}}
\nonumber\\
&&\theta(n-1) \* \biggl\{
      {1 \over n+1} \* {(1 - x_1)^{n+1} \over x_1} \* \biggl[
      S(n+1;1,1;(1-x_1\*x_2)/(1 - x_1),1/(1 - x_1\*x_2))
\nonumber\\&&
    - S(n+1;1,1;(1 - x_1\*x_2)/(1 - x_1),(1 - x_1\*x_2\*x_3)/(1 - x_1\*x_2)) \biggr]
    - x_2 \* x_3
   \biggr\} \, .
\end{eqnarray}
Upon replacement $n \to j_1$, the right-hand side of Eq.~(\ref{eq:demoresult}) 
could serve as input to another 
summation, thus nicely illustrating the telescoping character of the 
recursions discussed in Section~\ref{sec:harmonicsums}. 

In passing, we note that in some cases finite polynomial sums appear.
For example we may encounter sums of the types 
\begin{displaymath}
   \sum_{j=1}^n x^j\*j^k,\quad   \sum_{j=1}^n j^k,\quad  
   \sum_{j=1}^{n-1}x^j\*j^k,\quad   
   \sum_{j=1}^{n-1} j^k.   
\end{displaymath}
Internally they are called \keyword{xpowsum}, \keyword{powsum},
\keyword{xpowsum1}, \keyword{powsum1}. 
For positive integer values of $k$ up to 10 these sums are tabulated in the 
file \keyword{declsums.h}.
Generally, for any integer $k$, these types of sums can easily be obtained 
with programs such as \Maple\ or \Mathematica, 
where the corresponding recursive algorithms are implemented, 
see e.g. Ref.~\cite{Knuth:ConcreteMath}.

\subsection{List of procedures provided by \XSummer}

As briefly mentioned above, we distinguish two sets of procedures:
those easily accessible to the user and those for internal use only.
\begin{figure}[htbp]
  \begin{center}
    \leavevmode
    \includegraphics[width=0.95\textwidth]{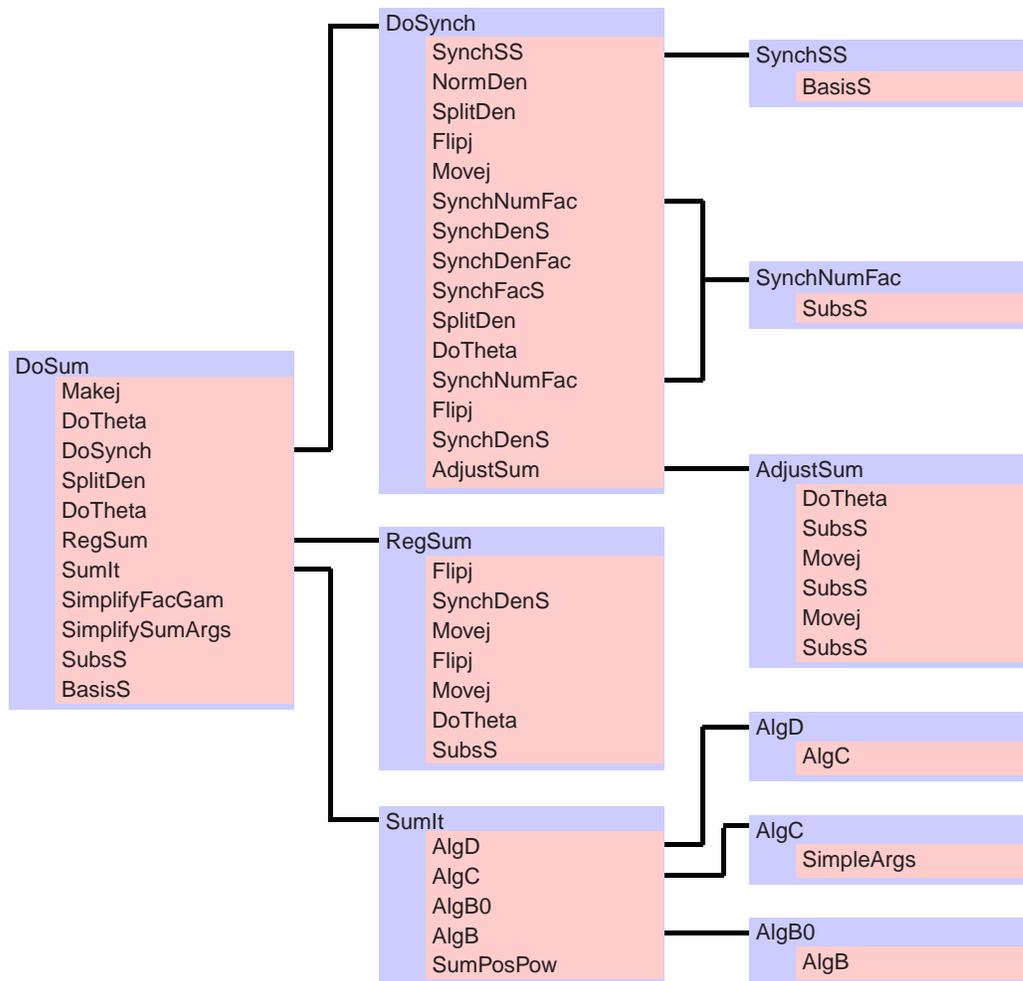}    
    \caption{Internal structure of the \XSummer\ package }
    \label{fig:InternalStructure}
  \end{center}
\end{figure}

\subsubsection{Procedures to be called directly}
\def\description#1{\parbox{\textwidth}{{\bfseries\sffamily #1}\\[0.2cm]
\hspace*{0.05\textwidth}\parbox{0.9\textwidth}{
     \input{./tmp/#1}}}\\[0.4cm]}
\begin{minipage}[h]{\textwidth}
{\bfseries\sffamily BasisS}\\[0.2cm]
\hspace*{0.05\textwidth}\begin{minipage}[h]{0.9\textwidth}
Express products of $S$-sums into single $S$-sums of higher weight.
\end{minipage}
\end{minipage}
\\[0.4cm]
\begin{minipage}[h]{\textwidth}
{\bfseries\sffamily ConvStoZ}\\[0.2cm]
\hspace*{0.05\textwidth}\begin{minipage}[h]{0.9\textwidth}
Convert $S$-sums into $Z$-sums.
\end{minipage}
\end{minipage}
\\[0.4cm]
\begin{minipage}[h]{\textwidth}
{\bfseries\sffamily DoSum}\\[0.2cm]
\hspace*{0.05\textwidth}\begin{minipage}[h]{0.9\textwidth}
User front-end to the \XSummer\ package. The first parameter
denotes the index of the innermost summation, the 
second is the index of the outermost sum, which should be summed.
For example
\begin{verbatim}
    #call DoSum(3,1)
\end{verbatim}
would do the sums over $j_3$, $j_2$, $j_1$ in this order.
\end{minipage}
\end{minipage}
\\[0.4cm]

An overview of the internal structure of \keyword{DoSum} is shown in 
Fig.~\ref{fig:InternalStructure}.

\subsubsection{Internal procedures}
\begin{minipage}[h]{\textwidth}
{\bfseries\sffamily AdjustSum}\\[0.2cm]
\hspace*{0.05\textwidth}\begin{minipage}[h]{0.9\textwidth}
Adjusts the sum boundaries. The argument specifies 
the index $i$ of the sum to be adjusted.
\end{minipage}
\end{minipage}
\\[0.4cm]
\begin{minipage}[h]{\textwidth}
{\bfseries\sffamily AlgB}\\[0.2cm]
\hspace*{0.05\textwidth}\begin{minipage}[h]{0.9\textwidth}
Implementation of the convolution algorithm (type B) to perform sums 
of the type
\begin{displaymath}
    \sum\limits_{i=1}^{n-1} 
    {x_1^i \over i^{m_1}} S(i;m_2,...,m_k;x_2,...,x_k) 
    {(x^{\prime}_1)^{n-i} \over (n-i)^{m^{\prime}_1}} 
    S(n-i;m^{\prime}_2,...,m^{\prime}_l;
									x^{\prime}_2,...,x^{\prime}_l)\, ,
\end{displaymath}
for any (integer) values of $m_i, m_i^\prime$ and  $m_1>0$.
The argument specifies 
the index $i$ of the sum to be done.
The special case $m_1=0$ is handled in \keyword{AlgB0}.

\end{minipage}
\end{minipage}
\\[0.4cm]
\begin{minipage}[h]{\textwidth}
{\bfseries\sffamily AlgB0}\\[0.2cm]
\hspace*{0.05\textwidth}\begin{minipage}[h]{0.9\textwidth}
Special case of the convolution algorithm (type B), 
for details see \keyword{AlgB}.
\end{minipage}
\end{minipage}
\\[0.4cm]
\begin{minipage}[h]{\textwidth}
{\bfseries\sffamily AlgC}\\[0.2cm]
\hspace*{0.05\textwidth}\begin{minipage}[h]{0.9\textwidth}
Implementation of the conjugation algorithm (type C) to perform sums 
of the form
\begin{eqnarray*}
   \sum\limits_{i=1}^{n}\, 
     \left( \begin{array}{c} n \\ i \end{array} \right)\, 
       (-1)^i {x_1^i \over i^{m_1}} 
       S(i;m_2,...,m_k;x_2,...,x_k)\, 
\end{eqnarray*}
for any (integer) values of $m_i$ and $m_1>0$. 
The argument specifies 
the index $i$ of the sum to be done. 
The special case $m_1=0$ is treated in \keyword{AlgC0}.
\end{minipage}
\end{minipage}
\\[0.4cm]
\begin{minipage}[h]{\textwidth}
{\bfseries\sffamily AlgC0}\\[0.2cm]
\hspace*{0.05\textwidth}\begin{minipage}[h]{0.9\textwidth}
Special case of the conjugation algorithm (type C), 
for details see \keyword{AlgC}.
\end{minipage}
\end{minipage}
\\[0.4cm]
\begin{minipage}[h]{\textwidth}
{\bfseries\sffamily AlgD}\\[0.2cm]
\hspace*{0.05\textwidth}\begin{minipage}[h]{0.9\textwidth}
Implementation of the binomial convolution algorithm (type D) 
to perform sums of the form
\begin{eqnarray*}
    &&\sum\limits_{i=1}^{n-1}\, 
    \left(\begin{array}{c} n \\ i \end{array} \right)\, 
     (-1)^i {x_1^i \over i^{m_1}} 
      S(i;m_2,...,m_k;x_2,...,x_k)\\
    &&\hspace{1cm}\times 
    {(x^\prime_1)^{n-i} \over (n-i)^{m^{\prime}_1}} 
     S(n-i;m^{\prime}_2,...,m^{\prime}_l;
          x^{\prime}_2,...,x^{\prime}_l)\, 
\end{eqnarray*}
for any (integer) values of $m_i,m_i^\prime \ge 0$ and $m_1>0$.
The argument specifies 
the index $i$ of the sum to be done. 
The special case $m_1=0$ is treated in \keyword{AlgD0}.
\end{minipage}
\end{minipage}
\\[0.4cm]
\begin{minipage}[h]{\textwidth}
{\bfseries\sffamily AlgD0}\\[0.2cm]
\hspace*{0.05\textwidth}\begin{minipage}[h]{0.9\textwidth}
Special case of the binomial convolution algorithm (type D); 
see \keyword{AlgD}.

\end{minipage}
\end{minipage}
\\[0.4cm]
\begin{minipage}[h]{\textwidth}
{\bfseries\sffamily DoSynch}\\[0.2cm]
\hspace*{0.05\textwidth}\begin{minipage}[h]{0.9\textwidth}
Procedure to synchronize the arguments of 
\keyword{num}, \keyword{den}, \keyword{fac}, \keyword{invfac} 
and \keyword{S}. The argument specifies 
the index $i$ of the sum to be adjusted.
Start with synchronizing products of \keyword{S} functions, 
then do the combinations 
\begin{enumerate}
\item $j_i$ and \keyword{fac}, \keyword{invfac}
\item \keyword{den} and \keyword{S}
\item \keyword{den} and \keyword{fac}, \keyword{invfac}
\item \keyword{S} and \keyword{fac}, \keyword{invfac}.
\end{enumerate}
Finally summation boundaries are synchronized.
\end{minipage}
\end{minipage}
\\[0.4cm]
\begin{minipage}[h]{\textwidth}
{\bfseries\sffamily DoTheta}\\[0.2cm]
\hspace*{0.05\textwidth}\begin{minipage}[h]{0.9\textwidth}
Simplifies combinations of 
\keyword{theta}, \keyword{delta}, \keyword{deltap}
and \keyword{sum} functions, defined for each sum over $j_i$. 
The argument specifies the index $i$ of the sum to be adjusted 
(argument $i=0$ implies no sum).
\end{minipage}
\end{minipage}
\\[0.4cm]
\begin{minipage}[h]{\textwidth}
{\bfseries\sffamily ExpandDen}\\[0.2cm]
\hspace*{0.05\textwidth}\begin{minipage}[h]{0.9\textwidth}
 Expands \keyword{den} function in small parameter \keyword{ep}, 
 i.e. $1/(a+b\epsilon)$ in terms of $\epsilon$. 
 If the argument provided to \keyword{ExpandDen}
 is 0 then denominators with integer values of $a$ are expanded.
 If the argument is 1, we expand for symbolic $a$.
\end{minipage}
\end{minipage}
\\[0.4cm]
\begin{minipage}[h]{\textwidth}
{\bfseries\sffamily ExpandGam}\\[0.2cm]
\hspace*{0.05\textwidth}\begin{minipage}[h]{0.9\textwidth}
Expands \keyword{Gamma} and \keyword{InvGamma} functions 
in the small parameter \keyword{ep}, 
i.e. $\Gamma(i+a\epsilon)$ and 
$1/\Gamma(i+a\epsilon)$ in $\epsilon$, where $i$ and $a$ take 
integer values.
The argument specifies the number of the highest sum 
in the expression. 
We choose the \mbox{$\overline{\mbox{MS}}$-scheme}, i.e. 
$\exp(-\gamma_E a \epsilon) = 1$, where $\gamma_E$ is 
Euler's constant.
\end{minipage}
\end{minipage}
\\[0.4cm]
\begin{minipage}[h]{\textwidth}
{\bfseries\sffamily Flipj}\\[0.2cm]
\hspace*{0.05\textwidth}\begin{minipage}[h]{0.9\textwidth}
Reverses the direction of summation. The argument specifies 
the index $i$ of the sum to be flipped.
\end{minipage}
\end{minipage}
\\[0.4cm]
\begin{minipage}[h]{\textwidth}
{\bfseries\sffamily Makej}\\[0.2cm]
\hspace*{0.05\textwidth}\begin{minipage}[h]{0.9\textwidth}
Creates the {\it dressed objects}, for example the \keyword{pow}
function {\tt pow(x,j3)} is converted to {\tt pow3(x,j3)}. 
The argument specifies the summation index to be treated.
\end{minipage}
\end{minipage}
\\[0.4cm]
\begin{minipage}[h]{\textwidth}
{\bfseries\sffamily Movej}\\[0.2cm]
\hspace*{0.05\textwidth}\begin{minipage}[h]{0.9\textwidth}
Makes a translation of $j_i+n$ $\to$ $j_i$. The first argument
specifies which index $i$ of the summation variable which should be 
used. The shift is determined by the argument of the function 
specified by the second argument passed to \keyword{Movej}. 
\end{minipage}
\end{minipage}
\\[0.4cm]
\begin{minipage}[h]{\textwidth}
{\bfseries\sffamily NormDen}\\[0.2cm]
\hspace*{0.05\textwidth}\begin{minipage}[h]{0.9\textwidth}
Converts \keyword{den} functions to normal form, i.e. fixes 
the sign of the summation index appearing in the 
denominator. The argument specifies the highest summation index 
$i$ to be treated. The hierarchy is such that $j_1$ is more 
important than $j_2$ etc. 
\end{minipage}
\end{minipage}
\\[0.4cm]
\begin{minipage}[h]{\textwidth}
{\bfseries\sffamily NormalizeGam}\\[0.2cm]
\hspace*{0.05\textwidth}\begin{minipage}[h]{0.9\textwidth}
Normalizes products of \keyword{Gamma} and \keyword{InvGamma} 
functions.
\end{minipage}
\end{minipage}
\\[0.4cm]
\begin{minipage}[h]{\textwidth}
{\bfseries\sffamily RegSum}\\[0.2cm]
\hspace*{0.05\textwidth}\begin{minipage}[h]{0.9\textwidth}
Performs final regularization of a sum. The argument specifies 
the index $i$ of the sum to be regularized.
\end{minipage}
\end{minipage}
\\[0.4cm]
\begin{minipage}[h]{\textwidth}
{\bfseries\sffamily SimpleArgs}\\[0.2cm]
\hspace*{0.05\textwidth}\begin{minipage}[h]{0.9\textwidth}
Applies some trivial simplifications. If the argument passed
to \keyword{SimpleArgs} is 0 these simplifications are applied to
the prefactors multiplying functions. If the argument passed 
is C the simplifications are applied to the arguments of
\keyword{S} and \keyword{pow} functions. 
\end{minipage}
\end{minipage}
\\[0.4cm]
\begin{minipage}[h]{\textwidth}
{\bfseries\sffamily Simplify}\\[0.2cm]
\hspace*{0.05\textwidth}\begin{minipage}[h]{0.9\textwidth}
Calls different simplification and normalization routines to
simplify or normalize the expresssions.
\end{minipage}
\end{minipage}
\\[0.4cm]
\begin{minipage}[h]{\textwidth}
{\bfseries\sffamily SimplifyFacGam}\\[0.2cm]
\hspace*{0.05\textwidth}\begin{minipage}[h]{0.9\textwidth}
Procedure to simplify  products of \keyword{fac}, \keyword{invfac} 
and \keyword{Gamma}, \keyword{InvGamma} functions.
\end{minipage}
\end{minipage}
\\[0.4cm]
\begin{minipage}[h]{\textwidth}
{\bfseries\sffamily SimplifySumArgs}\\[0.2cm]
\hspace*{0.05\textwidth}\begin{minipage}[h]{0.9\textwidth}
Procedure to simplify polynomials in arguments of 
\keyword{S} functions. This procedure is user accessible 
and should be edited if optimization is needed. 
\end{minipage}
\end{minipage}
\\[0.4cm]
\begin{minipage}[h]{\textwidth}
{\bfseries\sffamily SplitDen}\\[0.2cm]
\hspace*{0.05\textwidth}\begin{minipage}[h]{0.9\textwidth}
Partial fractioning of products of denominators involving
the summation variable $j_i$, where $i$ is the argument passed
to \keyword{SplitDen}. 
This routine is optimized for higher powers of the denominators 
and uses the sum formula rather than repeated splitting of pairs. 
\end{minipage}
\end{minipage}
\\[0.4cm]
\begin{minipage}[h]{\textwidth}
{\bfseries\sffamily SubsS}\\[0.2cm]
\hspace*{0.05\textwidth}\begin{minipage}[h]{0.9\textwidth}
Evaluates \keyword{S} functions with a numerical last argument 
into polynomials in the variables (if possible, with numerical 
values). For argument $i=0$ only numerical values are realized;
for $i=1$ the procedure also expands polynomials. 
\keyword{S} functions with symbolic last argument are untouched, 
of course. 
\end{minipage}
\end{minipage}
\\[0.4cm]
\begin{minipage}[h]{\textwidth}
{\bfseries\sffamily SumIt}\\[0.2cm]
\hspace*{0.05\textwidth}\begin{minipage}[h]{0.9\textwidth}
Calls the various procedures for summation algorithms. The argument 
specifies the index $i$ of the sum to be done.
\end{minipage}
\end{minipage}
\\[0.4cm]
\begin{minipage}[h]{\textwidth}
{\bfseries\sffamily SumPosPow}\\[0.2cm]
\hspace*{0.05\textwidth}\begin{minipage}[h]{0.9\textwidth}
Sums positive (or zero) powers of $j_i$ possibly in combination
with one \keyword{S} or \keyword{pow} function. The argument 
specifies the index $i$ of the sum to be done.
\end{minipage}
\end{minipage}
\\[0.4cm]
\begin{minipage}[h]{\textwidth}
{\bfseries\sffamily SynchDenFac}\\[0.2cm]
\hspace*{0.05\textwidth}\begin{minipage}[h]{0.9\textwidth}
Synchronizes combinations of \keyword{den} and \keyword{fac}, 
\keyword{invfac} functions, i.e. factorials and denominators. 
The argument specifies the associated summation index $i$.
\end{minipage}
\end{minipage}
\\[0.4cm]
\begin{minipage}[h]{\textwidth}
{\bfseries\sffamily SynchDenS}\\[0.2cm]
\hspace*{0.05\textwidth}\begin{minipage}[h]{0.9\textwidth}
Synchronizes combinations of \keyword{den} and \keyword{S}  
functions, i.e. denominators and $S$-sums.
The argument specifies the associated summation index $i$.
\end{minipage}
\end{minipage}
\\[0.4cm]
\begin{minipage}[h]{\textwidth}
{\bfseries\sffamily SynchFacS}\\[0.2cm]
\hspace*{0.05\textwidth}\begin{minipage}[h]{0.9\textwidth}
Synchronizes combinations of \keyword{S} and \keyword{fac}, 
\keyword{invfac} functions, i.e. factorials and $S$-sums. 
The argument specifies the associated summation index $i$.

\end{minipage}
\end{minipage}
\\[0.4cm]
\begin{minipage}[h]{\textwidth}
{\bfseries\sffamily SynchNumFac}\\[0.2cm]
\hspace*{0.05\textwidth}\begin{minipage}[h]{0.9\textwidth}
Synchronizes combinations of (positive) powers of $j_i$ and 
\keyword{fac}, \keyword{invfac} functions.
The argument specifies the associated summation index $i$.
\end{minipage}
\end{minipage}
\\[0.4cm]
\begin{minipage}[h]{\textwidth}
{\bfseries\sffamily SynchSS}\\[0.2cm]
\hspace*{0.05\textwidth}\begin{minipage}[h]{0.9\textwidth}
Synchronizes products of $S$-sums.
The argument specifies the associated summation index $i$.
\end{minipage}
\end{minipage}
\\[0.4cm]

As discussed in Section~\ref{sec:harmonicsums} the $S$-sums can be
treated as generalization of the harmonic sums investigated in 
Refs.~\cite{Euler:1775xx,Vermaseren:1998uu,Blumlein:2003gb}. 
It is therefore evident that some of the procedures 
presented here are similar to those in the \Summer\ package~\cite{Vermaseren:1998uu} written
by J.A.M.~Vermaseren.
In particular the procedures \keyword{AdjustSum}, \keyword{BasisS},
\keyword{DoSynch}, \keyword{DoTheta}, \keyword{Flipj}, \keyword{Makej}, 
\keyword{SubsS}, \keyword{SynchDenFac} and \keyword{SynchSS} 
are adapted versions of similar procedures in \Summer.
 
\subsubsection{Harmonic sums in infinity}

A special class of sums, occurring as a result of the summation algorithms, 
are harmonic sums in infinity. 
These are related to multiple zeta values~\cite{Hoffman:refs}, which for a
given weight are reducible to a small set of transcendental numbers 
using, for example, the algebraic properties in Eq.~(\ref{eq:prod-S}).

Together with the \XSummer\ package, we provide the tables of harmonic
sums in infinity (limited to weight six). 
These procedures are called \keyword{tables}, \keyword{table1}, ... , \keyword{table6} 
and the corresponding files are directly extracted from the \Summer\ 
package~\cite{Vermaseren:1998uu}. 
This facilitates interfacing with routines of the current package.

Let us also note that the reduction of multiple zeta values of a given weight 
to some irreducible set of constants is currently an active field of mathematical research.
Currently, downloads of tables up to weight nine in \Form\ \cite{Vermaseren:MZVtabs} 
or \Maple\ format~\cite{Petitot:MZVtabs} are publicly
available and extensions 
are known up to weight 16~\cite{Petitot:MZVtabs16,Vermaseren:MZVtabs16}.

%
% ---------------------------------------------------------------------
%
\setcounter{equation}{0}
\section{Examples}
\label{sec:examples}
%
% ---------------------------------------------------------------------
%

Along with the distribution of the \XSummer\  package, we provide also a
number of non-trivial examples.
These examples can either be run with the help of the shell script \keyword{TestIt} 
in a standard Linux/Unix environment or 
with the help of \keyword{TestItXP.bat} under the Microsoft Windows XP 
operating system. 
(For either of the scripts the user might have to make small adaptations, though.)

The respective script 
executes the \Form\  files \keyword{Examples.frm} and \keyword{DoIntegrals.frm} 
(discussed in detail below) and performs a check on the output of the former computations 
against tabulated results with the help of the file \keyword{CheckResults.frm}.
In this way, the user can verify the correctness of the installation of the \XSummer\  package.
At the same time, the examples are meant to illustrate the usage of sums with the 
\XSummer\  package. 
In particular, we want to clarify the conventions for the input to the procedures 
of Section~\ref{sec:xsummer}.

We provide in the file \keyword{Examples.frm} a number of 
(generalized) hypergeometric functions from the original Ref.~\cite{Moch:2001zr},
\begin{verbatim}
       hypergeom2F1(a*ep,b*ep,1-c*ep,x1)
       hypergeom2F1(1,-ep,1-ep,x1)
       hypergeom3F2(-2*ep,-2*ep,1-ep,1-2*ep,1-2*ep,x1)
       appel2(1,1,ep,1+ep,1-ep,x1,x2)
\end{verbatim}
where the coefficients of Laurent series in $\epsilon$ are calculated up to a
given order in terms of multiple polylogarithms. 
The first example {\tt hypergeom2F1(a*ep,b*ep,1-c*ep,x1)} was actually given in Eq.~(\ref{eq:ex-2F1}).
In \keyword{Examples.frm} it is realized by the following set of substitutions:
\begin{verbatim}
id hypergeom2F1(a?,b?,c?,x1?) = sum1(j1,0,inf) * 
      Po(j1+a,a)*Po(j1+b,b)*InvPo(j1+c,c)*invfac(j1)* pow(x1,j1);

id    Po(x1?,x2?) = Gamma(x1)*InvGamma(x2);
id InvPo(x1?,x2?) = Gamma(x2)*InvGamma(x1);
\end{verbatim}
Here {\tt Po(x1,x2)} and {\tt InvPo(x1,x2)} denote the Pochhammer symbols, 
see Eq.~(\ref{eq:ex-2F1}).

Next, the file \keyword{DoIntegrals.frm} calculates certain one- and two-loop Feynman integrals 
used in a complete calculation of higher-order perturbative corrections in 
Quantum Chromodynamics~\cite{Moch:2002hm}. 
To that end, integrals of various topologies had to be considered.
The respective analytical representations in terms of nested sums valid for arbitrary powers of denominators 
for the so-called C-topology~\cite{Moch:2001zr}, the B-box (one-loop box with one external mass~\cite{Anastasiou:1999cx})
and one-loop triangles with up to two external masses~\cite{Anastasiou:1999ui} 
are all given in the procedure \keyword{Int2Sum}.
The input to \keyword{DoIntegrals.frm}, i.e. the information on the topology and on the particular values 
for the powers of denominators, is passed on by preprocessor variables in \Form. 
The explicit lists of all inputs are contained in \keyword{SelectIntegral}.

In the actual calculation, we choose dimensional regularization~\cite
{'tHooft:1972fi,Bollini:1972ui,Ashmore:1972uj,Cicuta:1972jf} 
with $D=4-2\epsilon$ 
and the modified~\cite{Bardeen:1978yd} minimal subtraction~\cite{'tHooft:1973mm} 
scheme for all loop integrals in this section.
For illustration we give here the explicit series representation as we use it in \keyword{Int2Sum} 
for the one-loop triangle with two external masses $\mbox{Tri}(m,\nu_1,\nu_2,\nu_3;x_1)$.
It is defined by
\begin{eqnarray}
\label{eq:tri-example}
     \mbox{Tri}(m,\nu_1,\nu_2,\nu_3;x_1) = 
         \left( -s_{12} \right)^{-m+\epsilon+\nu_{123}}
         \int \frac{d^D k_1}{i \pi^{D /2}}
         \frac{1}{\left(-k_1^2\right)^{\nu_1}}
         \frac{1}{\left(-k_2^2\right)^{\nu_2}}
         \frac{1}{\left(-k_3^2\right)^{\nu_3}} \, 
\end{eqnarray}
and,  in the following, we use the short-hand notation $\nu_{ij}=\nu_i+\nu_j$ for sums of powers of 
propagators.
We have $k_2=k_1-p_1$, $k_3=k_2-p_2$, and we define the quantities
\begin{eqnarray}
  s_{12} = \left( p_1 + p_2 \right)^2\, , \quad\quad\quad\quad 
  x = \frac{p_1^2}{s_{12}} \, .
\end{eqnarray}

Equation~(\ref{eq:tri-example}) can be written as a combination of hypergeometric functions 
$_2F_1$. The series representation for this integral with $|x| \le 1$ is given by
\begin{eqnarray}
\label{eq:trisum}
&&         \mbox{Tri}(m,\nu_1,\nu_2,\nu_3;x) \,= \,
             \frac{ 
                     \Gamma(\epsilon-m+\nu_{23})
                     \Gamma(1-\epsilon+m-\nu_{23})
                     \Gamma(m-\epsilon-\nu_{13})
                  }{
                     \Gamma(\nu_1) \Gamma(\nu_2) \Gamma(\nu_3) 
                     \Gamma(2m-2\epsilon-\nu_{123})
                   } 
\nonumber \\ & & 
  \hspace{0.5cm}           \times 
             \sum\limits_{i=0}^\infty
                 \frac{x^{i}}{i!} 
             \bigg[
             x^{m-\epsilon-\nu_{23}} 
             \frac{
                     \Gamma(i_1+\nu_1) 
                     \Gamma(i_1-\epsilon+m-\nu_2) 
                  }{
                     \Gamma(i_1+1+m-\epsilon-\nu_{23}) 
                   }
\nonumber \\ && \hspace*{2.5cm}
                 - 
             \frac{
                     \Gamma(i_1+\nu_3) 
                     \Gamma(i_1-m+\epsilon+\nu_{123})
                  }{
                     \Gamma(i_1+1-m+\epsilon+\nu_{23}) 
                   }
             \bigg]\, .
\end{eqnarray}
Equation~(\ref{eq:trisum}) is implemented in the procedure \keyword{Int2Sum} and prepared 
for the input to the \XSummer\  package with a set of substitutions similar to those briefly 
discussed above for the $_2F_1$ hypergeometric function.

Now, with the examples at hand, we would like to discuss the efficiency of the 
implementation of the \XSummer\  package in \Form.
As already mentioned above, certainly one major disadvantage in using \Form\ 
is the lack of internal algorithms for polynomial algebra.
As a consequence, we cannot easily bring rational functions of polynomials to a normal form. 
For that purpose, we would need operations such as factorization, polynomial division 
or combinations thereof as used for instance in partial fractioning.
In applications of the algorithms A--D to sums with multiple scales, this problem becomes apparent 
when trying to normalize the arguments of the $S$-sums.
In particular, the recursions of the conjugation algorithm (type C) are sensitive to this issue.
Here, efficient simplifications of polynomials can have a significant impact on the execution time 
for a given sum.

Since no really elegant way for simplifying polynomials exists within \Form, 
we have provided a case-by-case solution.
Inside the procedure \keyword{AlgC} we do have the (user-accessible) procedure \keyword{SimpleArgs} 
(described in the previous section). 
There, depending on the scales involved, appropriate substitutions for tuning or optimizations should be added.
Then, we do have the procedure \keyword{CustomizeDen}. It uses partial fractioning and normalizes 
the ubiquitous denominators to standard factors.
In a similar spirit, we also use the procedure \keyword{ParFrac}, 
which brings two-scale polynomials to normal form in \keyword{DoIntegrals.frm} by means of partial fractioning 
at the end of the calculation.

Another more generic approach to this problem, which the experienced user might consider, is the following.
\Form\  provides the preprocessor statements \keyword{\#system} and \keyword{\#pipe}, which invoke 
a call to the operating system.
This provides the opportunity to realize a link to external programs, in particular to 
computer algebra systems such as \Maple\  or \Mathematica\  along with their functionalities in  
polynomial algebra.
However, both the detailed description and the implementation of these features is beyond the scope 
of the present work.

Finally, we would like to give some information on the use of computer resources.
Typically, the execution times and the use of memory or disk space depend very much on the problem 
under consideration.
As a general rule, the expression size and therefore the execution time correlate strongly with 
the depth of the Laurent expansion in $\epsilon$. In the examples, we can control this with the 
preprocessor variable {\tt EXPANDEP}. This effectively cuts the series expansion in $\epsilon$ at 
the specified power. Note that for every individual term the series is cut 
to the specified order. If additional poles are present it might thus 
be necessary to expand individual terms to higher order to make sure that
no terms are lost. 
For safeguard, we added a function {\tt order}, which shows 
the effect of the truncation.
Of course, the run time is also correlated with the number of nested sums.
In practical applications it might be  advantageous to start with a low
value for {\tt EXPANDEP} and increase it until the final result has the desired
order.
In addition to this, in practical applications, some tuning may be needed for new large problems,
particularly in bringing polynomials to normal form.

Let us close this section with a few remarks about the runtime performance. 
To give the interested user a hint on how long the presented examples will 
take, we measured the runtime on a standard PC. 
In particular, the machine we used was equipped with a 3~GHz P4 
(hyperthreading support) with 512~MB 
of RAM. The considered problems fit completely into the RAM so that the 
hard-drive speed plays only a marginal r\^ole. The results are shown in 
Table~\ref{tab:Runtimes}, where we averaged always over 10 
independent runs.\renewcommand{\arraystretch}{1.1}
\begin{table}[htbp]
    \label{tab:Runtimes}
  \centering
  \begin{tabular}[htbp]{l|c|c|c|c|}
    &
    &
   \multicolumn{3}{|c|}{\keyword{DoIntegrals.frm}}	
	\\	
\Form\ 3.1 executable&\keyword{Examples.frm}&
        CTOPO        
	    &
        BBOX        
    &
        TRIMASS        
    \\ \hline
    gcc, 15-jul-2005& 7.9s & 20.9s & 119.3s & 19s\\
    icc, 24-jan-2003& 7.5s & 19.9s & 108.2s&17.4s \\
    gcc, 17-oct-02& 8.7s & 24.5s & 181.3s  & 27.6s\\
    %%WinXP&7.95&20.50&111.78&18.31		
  \end{tabular}
  \caption{Runtimes of the examples under Linux for 
	different \Form\ versions.} 
\end{table}
\renewcommand{\arraystretch}{1.}%%
The different \Form\ versions used are  available at 
{\tt http://www.nikhef.nl/\~{}form}.
We also tried a preliminary version of a native 
Microsoft Windows XP 
\Form\ executable.
The runtimes are very similar to those obtained with the icc-compiled \Form\ 
executable and the new gcc version. The difference of 5\% might be due to the fact that on 
Windows XP
the total time was measured and not only the user time.

In applications more involved than the examples shown here, 
the user is advised to make a detailed analysis 
on the actual depth of the expansion in $\epsilon$ needed and on 
the particular structure of the polynomials. Furthermore the user should 
provide additional routines to simplify the symbolic arguments of $S$-sums.

%
% ---------------------------------------------------------------------
%
\setcounter{equation}{0}
\section{Conclusion}
\label{sec:conclusion}
%
% ---------------------------------------------------------------------
%

Symbolic summation has advanced to an important method, e.g. in perturbative quantum field theory, 
and significant progress has been made during the past years.
In the present work, we have provided an implementation in \Form\ 
of algorithms suitable for the expansion of transcendental functions 
in a small parameter around integer values, such that the resulting 
(generalized) hypergeometric series can be expressed in closed form in multiple 
polylogarithms.

As examples, we have discussed various applications in quantum field theory, particularly 
in the calculation of Feynman diagrams at higher orders in perturbation theory.
In this context, the \XSummer\ package has been used in perturbative calculations extensively 
and we believe it may be useful for a larger community.
We have chosen \Form\ for the implementation, because it is a fast and efficient 
computer algebra system and because of its capability to handle 
large expressions.
For convenience of the user, we provide along with \XSummer\ a set of 
sample calculations. These illustrate the use of the program.
An extension of the present implementation to cover also algorithms for generalized sums 
from expansions around rational numbers (see e.g.~\cite{Weinzierl:2004bn}) 
will be the subject 
of a future publication.

%
% ---------------------------------------------------------------------
%
\subsection*{Note added}
Very recently, Ref.~\cite{Huber:2005yg} appeared, which addresses the problem of expanding 
hypergeometric functions $_JF_{J-1}$ around integer parameters to arbitrary order.
It provides an implementation in \Mathematica\ of the algorithms (\ref{eq:type-A}), (\ref{eq:type-B}), 
i.e. type A and B of the original Ref.~\cite{Moch:2001zr}.
Thus, it is capable of performing some expansions already discussed in Ref.~\cite{Moch:2001zr} and discussed also in the present Letter, 
for instance Eq.~(\ref{eq:ex-2F1}).

%
% ---------------------------------------------------------------------
%

%
% ---------------------------------------------------------------------
%
\subsection*{Acknowledgements}
We would like to thank J.A.M~Vermaseren for his kind permission 
to use  some files of the \Summer\  package~\cite{Vermaseren:1998uu} 
in the current distribution.
The work of S.M. has been supported in part by 
the Helmholtz Gemeinschaft under contract VH-NG-105 and 
by the Deutsche Forschungsgemeinschaft in Sonderforschungsbereich/Transregio 9.
%
% ---------------------------------------------------------------------
%

\end{document}